\documentclass[preprint,twocolumn,sort&compress,3p,times]{elsarticle}

\usepackage{atlasphysics}

\usepackage{amsmath,graphicx}
\usepackage{a4wide}

\usepackage{booktabs}
\usepackage{numcompress}
\usepackage{lineno}
\usepackage{multirow,xspace}
\usepackage{rotating}
\usepackage{afterpage}
\usepackage[hyperindex,breaklinks]{hyperref} 

\graphicspath{{./figures/}}

\newcommand{\HWWlvqq}    {H\rightarrow WW\rightarrow \ell\nu jj}

\def\simge{\mathrel{%
   \rlap{\raise 0.511ex \hbox{$>$}}{\lower 0.511ex \hbox{$\sim$}}}}
\def\simle{\mathrel{
   \rlap{\raise 0.511ex \hbox{$<$}}{\lower 0.511ex \hbox{$\sim$}}}}

\usepackage{preprintcover}  
\PreprintCoverPaperTitle{Search for the Higgs boson in the $H\rightarrow
  WW\rightarrow\ell\nu jj$ decay channel at $\boldmath{\sqrt{s}}$ =
  7\,TeV with the ATLAS detector}  
\PreprintIdNumber{CERN-PH-EP-2012-162}  
\PreprintCoverAbstract{
A search for the Standard Model Higgs boson has been performed in the
$H\rightarrow WW\rightarrow\ell\nu jj$ channel using 4.7 ${\rm fb}^{-1}$
of $pp$ collision data recorded at a centre-of-mass energy of
$\sqrt{s} = 7~{\rm TeV}$ with the ATLAS detector at the Large Hadron
Collider. Higgs boson candidates produced in association with zero,
one or two jets are included in the analysis to maximize the
acceptance for both gluon fusion and weak boson fusion Higgs boson
production processes. No significant excess of events is observed over
the expected background and limits on the Higgs boson production cross
section are derived for a Higgs boson mass in the range $300~{\rm
  GeV}<\mH<600$~GeV. The best sensitivity is reached for $\mH =
400\GeV$, where the observed (expected) 95\% confidence level upper
bound on the cross section for $H\rightarrow WW$ produced in
association with zero or one jet is 2.2 pb (1.9 pb), corresponding to
1.9 (1.6) times the Standard Model prediction. In the Higgs boson plus
two jets channel, which is more sensitive to the weak boson fusion
process, the observed (expected) 95\% confidence level upper bound on
the cross section for $H\rightarrow WW$ production with $\mH =
400\GeV$ is 0.7 pb (0.6 pb), corresponding to 7.9 (6.5) times the
Standard Model prediction.
}  
\PreprintJournalName{Physics Letters B}  

\journal{Physics Letters B}

\begin{document}


\begin{frontmatter}  

\title{Search for the Higgs boson in the $H\rightarrow
  WW\rightarrow\ell\nu jj$ decay channel at $\sqrt{s} = 7$
  TeV with the ATLAS detector}

\author{The ATLAS Collaboration}

\begin{abstract} 
A search for the Standard Model Higgs boson has been performed in the
$H\rightarrow WW\rightarrow\ell\nu jj$ channel using 4.7 ${\rm fb}^{-1}$
of $pp$ collision data recorded at a centre-of-mass energy of
$\sqrt{s} = 7~{\rm TeV}$ with the ATLAS detector at the Large Hadron
Collider. Higgs boson candidates produced in association with zero,
one or two jets are included in the analysis to maximize the
acceptance for both gluon fusion and weak boson fusion Higgs boson
production processes. No significant excess of events is observed over
the expected background and limits on the Higgs boson production cross
section are derived for a Higgs boson mass in the range $300~{\rm
  GeV}<\mH<600$~GeV. The best sensitivity is reached for $\mH =
400\GeV$, where the observed (expected) 95\% confidence level upper
bound on the cross section for $H\rightarrow WW$ produced in
association with zero or one jet is 2.2 pb (1.9 pb), corresponding to
1.9 (1.6) times the Standard Model prediction. In the Higgs boson plus
two jets channel, which is more sensitive to the weak boson fusion
process, the observed (expected) 95\% confidence level upper bound on
the cross section for $H\rightarrow WW$ production with $\mH =
400\GeV$ is 0.7 pb (0.6 pb), corresponding to 7.9 (6.5) times the
Standard Model prediction.

\end{abstract}


\end{frontmatter}  




\section{Introduction}

In the Standard Model (SM), a scalar field with a non-zero vacuum
expectation value breaks the electroweak symmetry, gives masses to the
$W/Z$ bosons and fermions~\cite{Englert:1964et,Higgs:1964ia,Higgs:1964pj,Guralnik:1964eu,Higgs:1966ev,Kibble:1967sv},
and manifests itself directly as a particle, the Higgs boson
\cite{Higgs:1964ia,Higgs:1964pj,Higgs:1966ev}. A primary goal of the
Large Hadron Collider (LHC) is to test the SM mechanism of electroweak
symmetry breaking by searching for Higgs boson production in
high-energy proton-proton collisions. At LHC energies, the Higgs boson
is predominantly produced via gluon fusion ($gg\rightarrow H$) and via
weak boson fusion ($qq\rightarrow qqH$).

Results of Higgs boson searches in various channels using data
up to an integrated luminosity of approximately 5 fb$^{-1}$ have
recently been reported by both the ATLAS and CMS
collaborations~\cite{ATLAS:2012ae,Chatrchyan:2012tx}. The ATLAS
analysis excludes a Higgs boson with mass in the ranges
112.9--115.5~GeV, 131--238~GeV and 251--466 GeV while the CMS analysis
excludes the range 127--600~GeV at 95\% confidence level (CL). Direct
searches at LEP and the Tevatron exclude Higgs boson masses $\mH
<114.4$ GeV~\cite{Barate:2003sz} and $156~\GeV < \mH <
177~\GeV$~\cite{tevhiggs86fb} respectively at 95\% CL.

For $m_H \simge 135~\GeV$, the dominant decay mode of the Higgs boson
is $H\rightarrow WW^{(\ast)}$. For $m_H\simge 200~\GeV$, the
$\HWWlvqq$ channel, where one $W$ boson decays into two quarks leading
to a pair of jets ($W\rightarrow jj$) and the other decays into a
charged lepton and a neutrino ($W\rightarrow\ell\nu$) where $\ell = e$
or $\mu$, becomes interesting since jets from the Higgs boson decay
are, on average, more energetic than the jets from the dominant
background ($W$+jets). An advantage of $\HWWlvqq$ over channels with
two final-state neutrinos is the possibility of reconstructing
the Higgs boson mass using kinematical constraints to estimate the
component of the neutrino momentum along the beam axis.

This Letter describes a search for the SM Higgs boson in the $\HWWlvqq$ 
channel using the ATLAS detector at the LHC, based on 4.7 fb$^{-1}$
of $pp$ collision data collected at a centre-of-mass energy $\sqrt{s}
= 7$ TeV during 2011. The present search supersedes a previous
analysis in the same Higgs boson decay channel published by the
ATLAS Collaboration~\cite{PhysRevLett.107.231801}. The distribution of
the $\ell\nu jj$ invariant mass $m(\ell\nu jj)$, reconstructed using
the $\ell\nu$ invariant mass constraint $m(\ell\nu) = m(W)$ and the
requirement that two of the jets in the event are consistent with a
$W\rightarrow jj$ decay, is used to search for a Higgs boson
signal. Feed-down from $\tau$ lepton decays is included in this
analysis for both background and signal, i.e. $H\to WW \to
\tau\bar{\nu}_\tau jj \to \ell\bar{\nu}_\ell\nu_\tau\bar{\nu}_\tau jj$.

The present search is restricted to $m_{H} > 300$ GeV in order to
ensure a smoothly varying non-resonant background. The search is
further limited to $m_{H} < 600~\GeV$ since, for higher Higgs boson
masses, the jets from $W\rightarrow jj$ decay begin to overlap due 
to the large boost of the $W$ boson, and the natural width of the
Higgs boson exceeds 100 GeV. The best sensitivity to Higgs boson
production in this analysis is expected for $m_{H} \sim 400$ GeV.


\section{The ATLAS Detector}

The ATLAS experiment~\cite{atlas} uses a multipurpose particle
detector with forward-backward symmetric cylindrical
geometry\footnote{ATLAS uses a right-handed coordinate system with its 
  origin at the nominal interaction point (IP) in the centre of the
  detector and the $z$-axis coinciding with the axis of the beam
  pipe. The $x$-axis points from the IP to the centre of the LHC ring,
  and the $y$-axis points upward. Cylindrical coordinates ($r$,$\phi$)
  are used in the transverse plane, $\phi$ being the azimuthal angle
  around the beam pipe. The pseudorapidity is defined in terms of the
  polar angle $\theta$, measured with respect to the $z$-axis, 
  as $\eta = -\ln[\tan(\theta/2)]$.}
covering the pseudorapidity range $|\eta|<2.5$ for charged particles
and $|\eta|<4.9$ for jet measurements. The inner tracking detector
(ID) consists of a silicon pixel detector, a silicon microstrip
detector, and a transition radiation tracker. The ID is surrounded by
a thin superconducting solenoid providing a 2~T axial magnetic
field. The superconducting solenoid is surrounded by a
high-granularity liquid-argon (LAr) sampling electromagnetic (EM)
calorimeter. An iron/scintillator tile calorimeter provides hadronic
coverage in the central rapidity range. The end-cap and forward
regions are instrumented with LAr calorimeters for both
electromagnetic and hadronic measurements. The muon spectrometer
surrounds the calorimeters and consists of three large superconducting
toroids, each with eight coils, a system of precision tracking
chambers, and detectors for triggering.


\section{Data and simulation samples}

The data were collected using single-muon and single-electron
triggers~\cite{atlas:trigger}. The single-muon trigger required the
transverse momentum ($\pt$) of the muon with respect to the beam line
to exceed 18 GeV; for the single-electron trigger, the threshold
varied from 20 GeV to 22 GeV. The trigger object quality requirements
were tightened throughout the data-taking period to cope with
increasing instantaneous luminosity. 

Using the ATLAS simulation framework~\cite{ATLAS_simulation_paper},
detailed Monte Carlo (MC) studies of signal and backgrounds have been 
performed. The interaction with the ATLAS detector is modelled with
GEANT4~\cite{geant4} and the events are processed through the same
reconstruction chain that is used to perform the reconstruction
of data events. The effect of multiple $pp$ interactions in the same
and nearby bunch crossings (pile-up) is modelled by superimposing
several simulated minimum-bias events on the simulated signal and
background events. Simulated MC events are weighted to match the
distribution of interactions per beam crossing in the dataset.


\section{Object Selection}

The $pp$ collision vertices in each bunch crossing are reconstructed
using the inner tracking system \cite{ATLAS-CONF-2010-069}. To remove
cosmic-ray and beam-induced backgrounds, events are required to have
at least one reconstructed primary vertex with at least three
associated tracks with $\pt>400$~MeV. If multiple collision vertices
are reconstructed, the vertex with the largest summed $p^2_T$ of the
associated tracks is selected as the primary vertex.

Each electron candidate is reconstructed from clustered energy deposits
in the EM calorimeter with an associated track. It is further required
to satisfy a tight set of identification criteria with an efficiency
of approximately 80\% for electrons from $W\to e\nu$ decays with
transverse energy $20~\GeV < \et < 50~\GeV$ \cite{Aad:2011mk}. While
the energy measurement is taken from the EM calorimeter, the
pseudorapidity $\eta$ and azimuthal angle $\phi$ are taken from the
associated track. The cluster is required to be in the range
$|\eta|<2.47$, excluding the transition region between barrel and
endcap calorimeters, $1.37<|\eta|<1.52$, and small calorimeter regions
affected by temporary operational problems. The track associated with
the electron candidate is required to point back to the reconstructed
primary vertex with a transverse impact parameter significance
$|d_{0}/\sigma_{d_{0}}|< 10$ and with an impact parameter along the
beam direction of $|z_{0}|< 1$~mm. Electrons are further required to
be isolated: the sum of the transverse energies (excluding the
electron itself) in calorimeter cells inside a cone $\Delta R \equiv
\sqrt{(\Delta\phi)^2+(\Delta\eta)^2}=0.3$ around the cluster
barycentre must satisfy
$\Sigma(E_{\mathrm{T}}^{\mathrm{calo}})/\pt^{e}<0.14$ and the scalar
sum of the transverse momenta of all tracks (excluding the electron
track itself) with $\pt>1$~GeV from the primary vertex in the same
cone must satisfy $\Sigma(\pt^{\mathrm{track}})/\pt^{e}<0.13$.

Muons are reconstructed by combining tracks in the inner detector and
the muon spectrometer. The identification efficiency is measured
to be $(92.8\pm 0.2)$\% for muons with transverse momentum $\pt>20$~GeV
\cite{ATLAS-CONF-2011-063}. Tracks are required to pass basic quality
cuts on the number and type of hits in the inner detector. They must
lie within the range $|\eta|<2.4$. The tracks must satisfy the same
$z_0$ cut as electrons and $|d_{0}/\sigma_{d_{0}}|< 3$. They must also
be isolated, with the sum of the transverse energies (excluding those
attributed to the muon itself) in calorimeter cells inside a cone
$\Delta R=0.3$ around the muon satisfying
$\Sigma(E_{\mathrm{T}}^{\mathrm{calo}})/\pt^{\mu}<0.14$. Furthermore,
the scalar sum of the transverse momenta of all tracks with
$\pt>1$~GeV from the primary vertex inside a cone $\Delta R=0.4$
around the muon must satisfy $\Sigma(\pt^{\mathrm{track}})/\pt^{\mu}<0.15$.

Jets are reconstructed from topological clusters of energy deposited 
in the calorimeters using the anti-$k_{t}$ algorithm~\cite{AntiKt}
with radius parameter $R=0.4$. The reconstructed jet energy is
calibrated using \pt- and $\eta$-dependent correction factors based on
MC simulation and validated with data~\cite{JES}. The selected jets
are required to have $\pt>25$~GeV and $|\eta|<4.5$. Jets are
considered $b$-tagged if they satisfy the requirement $|\eta|<2.8$ and
are consistent with having originated from the decay of a
$b$-quark. This latter requirement is determined by a $b$-tagging
algorithm which uses a combination of impact parameter significance
and secondary vertex information and exploits the topology of weak
decays of $b$- and $c$-hadrons. The algorithm is tuned to achieve an
80\% $b$-jet identification efficiency, which results in a tagging
rate for light quark jets of approximately
6\%~\cite{ATLAS-CONF-2012-040,ATLAS-CONF-2012-043}. The missing
transverse momentum and its magnitude \met\ are reconstructed from
calibrated jets, leptons and photons, and take into account soft
clustered energy in the calorimeters~\cite{MET}. Energy deposited by
muons is subtracted in the \met\ calculation to avoid double
counting.


\section{Event Selection}

Events are classified based on the number of jets selected in addition
to the two jets from the Higgs boson decay candidate. For events to be
selected as Higgs boson candidates without an additional jet ($H+0j$)
or with exactly one additional jet ($H+1j$), the channels which are
more sensitive to the gluon fusion process, the following conditions
must be met: only one reconstructed lepton candidate (electron or
muon) with $\pt>40$~GeV, no additional leptons with $\pt>20$~GeV,
$\MET > 40$~GeV, and exactly two jets ($\ell\nu jj+0$ jet sample) or
exactly three jets ($\ell\nu jj+1$ jet sample) with $\pt > 40$~GeV and
$|\eta|<4.5$. The two jets with invariant mass ($m_{jj}$) closest to
the mass of the $W$ boson are required to satisfy $71~\GeV < m_{jj} <
91~\GeV$. The more energetic of these two jets must satisfy $\pt >
60$~GeV. These two jets are taken as the $W$ boson decay jets and are
required to lie within the range $|\eta|<2.8$, where the jet energy
scale is best known (with an uncertainty of 5\% or less for
$\pt>40$~GeV, depending on $\pt$ and $|\eta|$ over this
range~\cite{JES}), and have $\Delta R_{jj}<1.3$ to suppress $W$+jets
background. In order to reduce top quark background, the event is
rejected if either of the $W$ boson decay jets is $b$-tagged.

For the $\ell\nu jj+2j$ selection ($H+2j$), which is more sensitive to
the weak boson fusion Higgs boson production mode, the following
requirements are applied. The charged lepton \pt\ and the \met\ must
both exceed 30~GeV. There must be at least four jets with $\pt>25$~GeV
and $|\eta|<4.5$. The two jets with invariant mass closest to the mass
of the $W$ boson are required to satisfy $71~\GeV < m_{jj} <
91~\GeV$. These jets are labelled as the $W$ boson decay jets. Because
of the small signal cross section in this channel, the $W$ boson decay
jets are not required to lie within $|\eta|<2.8$, in order to increase
the acceptance. The event is required to satisfy a set of ``forward
jet tagging" cuts designed to select $qq\rightarrow qqH$ events. The
two highest-\pt\ jets apart from the $W$ boson decay jets are labelled
as the ``tag" jets, and they are required to be in opposite hemispheres
($\eta_{j1}\cdot\eta_{j2}<0$). They are also required to be
well-separated in pseudorapidity
($\Delta\eta_{jj}=|\eta_{j1}-\eta_{j2}|>3$). The lepton is required to 
be between the two tag jets in pseudorapidity. The two tag jets must
have large invariant mass ($m_{jj}>600$~GeV) and there must be no
additional jets in the range $|\eta|<3.2$. The event is rejected if it
contains a $b$-tagged jet.

The $\ell\nu jj+0/1j$ selection differs from the selection used
Ref.~\cite{PhysRevLett.107.231801}. The selection criteria are
optimized to improve the expected Higgs boson sensitivity for masses
above 300~GeV and require a more complex parameterization
of the background shape, as discussed in Section~\ref{sec:modeling}.


\section{Expected Backgrounds}

In both the $\ell\nu jj+0/1j$ and $\ell\nu jj+2j$ selections, the
background is expected to be dominated by $W$+jets production. Other
important backgrounds are $Z$+jets, $t\bar{t}$, single top quark,
diboson ($WW$, $WZ$, $ZZ$, $W\gamma$ and $Z\gamma$) production, and
multijets (MJ) from strong interaction processes that can be selected
due either to the presence of leptons from heavy-flavour decays or 
jets misidentified as leptons.  

Although MC predictions are not used to model the background in the
Higgs boson search results, a combination of MC and data-driven
methods is used to understand the background composition at this
intermediate stage. Backgrounds due to $W/Z$+jets, $t\overline{t}$,
and diboson production are modelled using the ALPGEN~\cite{alpgen},
MC@NLO~\cite{mcatnlo}, and HERWIG~\cite{herwig} generators,
respectively. Single top production is modelled using
AcerMC~\cite{Kersevan:2004yg} and single top produced in association
with a $W$ boson is modelled with MC@NLO. The small contribution from
$W/Z+\gamma$ events is estimated from events simulated using
MadGraph/MadEvent~\cite{Alwall:2007st}. The shapes of MJ background
distributions are modelled using histograms derived from data samples
selected in the same way as for the $H\rightarrow WW\rightarrow\ell\nu
jj$ selection, except that the electron identification requirements
are loosened and the isolation requirement on muons is inverted. In
the loosened selection, electrons satisfying the complete set of
identification criteria are not included. Expected contributions from
top quark ($t\bar{t}$ and single top) production and electroweak boson
(including diboson) production to the MJ shape histograms are
subtracted using MC predictions.

To normalize the MJ background contribution, fits to the \met\ distribution
using templates for each background contribution are performed. The
\met\ template is constructed from the loose lepton control sample
after the selection is further relaxed by omitting the $\MET$
criteria. The normalizations of this MJ template and the corresponding
template for $W/Z$+jets taken from MC are fitted to the observed
$\MET$ distribution in data  after the final selection, with other
backgrounds estimated using the MC simulation and fixed to their
expectation for 4.7 fb$^{-1}$. The relative contributions from
$W$+jets and $Z$+jets into the $W/Z$+jets template are fixed according
to the SM cross sections. The scale factors for the MJ and $W/Z$+jets
templates derived from these fits are used to normalize the MJ and
$W/Z$+jets background contributions in comparisons between data and
these background expectations.

The MC simulation predicts that $W/Z$+jets events constitute
($72\pm14$)\% of the total background for $\ell\nu jj+0/1j$ and
($77\pm15$)\% for $\ell\nu jj+2j$, while the top quark backgrounds
contribute with ($19\pm5$)\% and ($9\pm2$)\% for $\ell\nu jj+0/1j$ and
$\ell\nu jj+2j$ respectively. 

\section{$WW$ Mass Reconstruction}

To reconstruct the invariant mass $m(\ell\nu jj)$ of the
$WW$ system, the neutrino momentum is required. Its transverse
momentum $\pt^{\nu}$ is taken from the measured $\MET$ while the
neutrino longitudinal momentum $p_z^\nu$ is computed using the second
degree equation given by the mass constraint $m(\ell\nu)=m(W)$. In the
case of two real solutions, the solution with smaller neutrino
longitudinal momentum $|p_{z}^{\nu}|$ is taken, based on simulation
studies. In the case of complex solutions, the event is rejected. This
requirement rejects ($20\pm1$)\% of MC signal events at
$m_{H}=400$~\GeV, while for MC $W$+jets the corresponding rejection is
($30\pm1$)\%. These estimates include only statistical
uncertainties. Larger fractions of events are rejected in $\ell\nu
jj+1j$ than in $\ell\nu jj+0j$ independent of lepton flavour. In
collision data ($30\pm1$)\% of the events are rejected by this
requirement, consistent with the expectations from the $W$+jets
background simulation.

\section{Signal and Background Modelling}\label{sec:modeling}

\begin{figure*}
\begin{center}
\includegraphics[angle=0,width=0.49\textwidth]{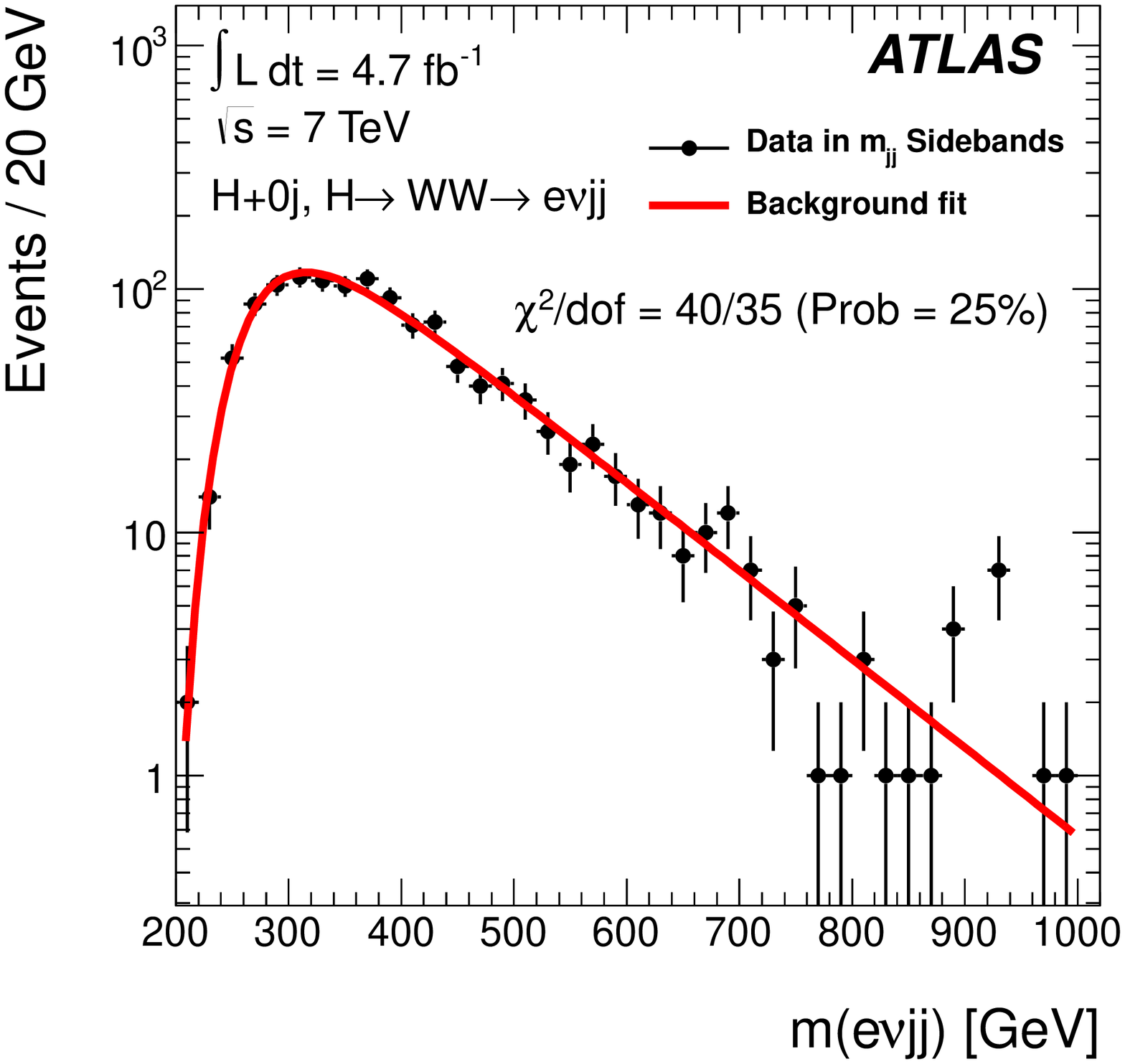} 
\includegraphics[angle=0,width=0.49\textwidth]{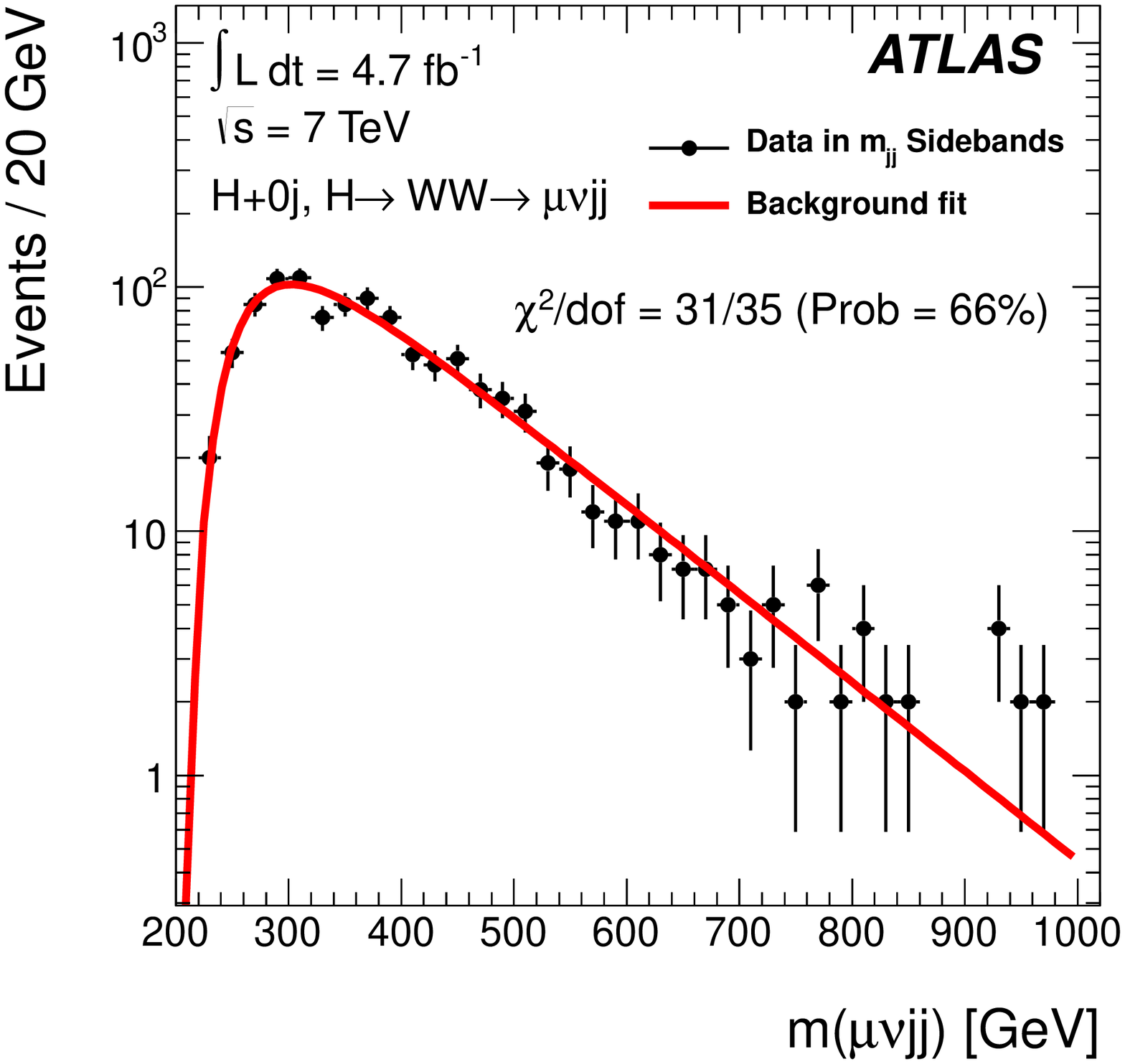}
\caption{Fits of the background model described in the text to the
  reconstructed invariant mass $m(\ell\nu jj)$ when $m_{jj}$ is in
  the $W$ sidebands for the $\ell\nu jj+0j$ selection. The left (right) figure
  shows the electron (muon) channel distribution. The $\chi^2$/dof
  and $\chi^2$ probability of these fits are also shown in the figure.
}
\label{sideband_fits_0j}
\end{center}
\end{figure*}

\begin{figure*}
\begin{center}
\includegraphics[angle=0,width=0.49\textwidth]{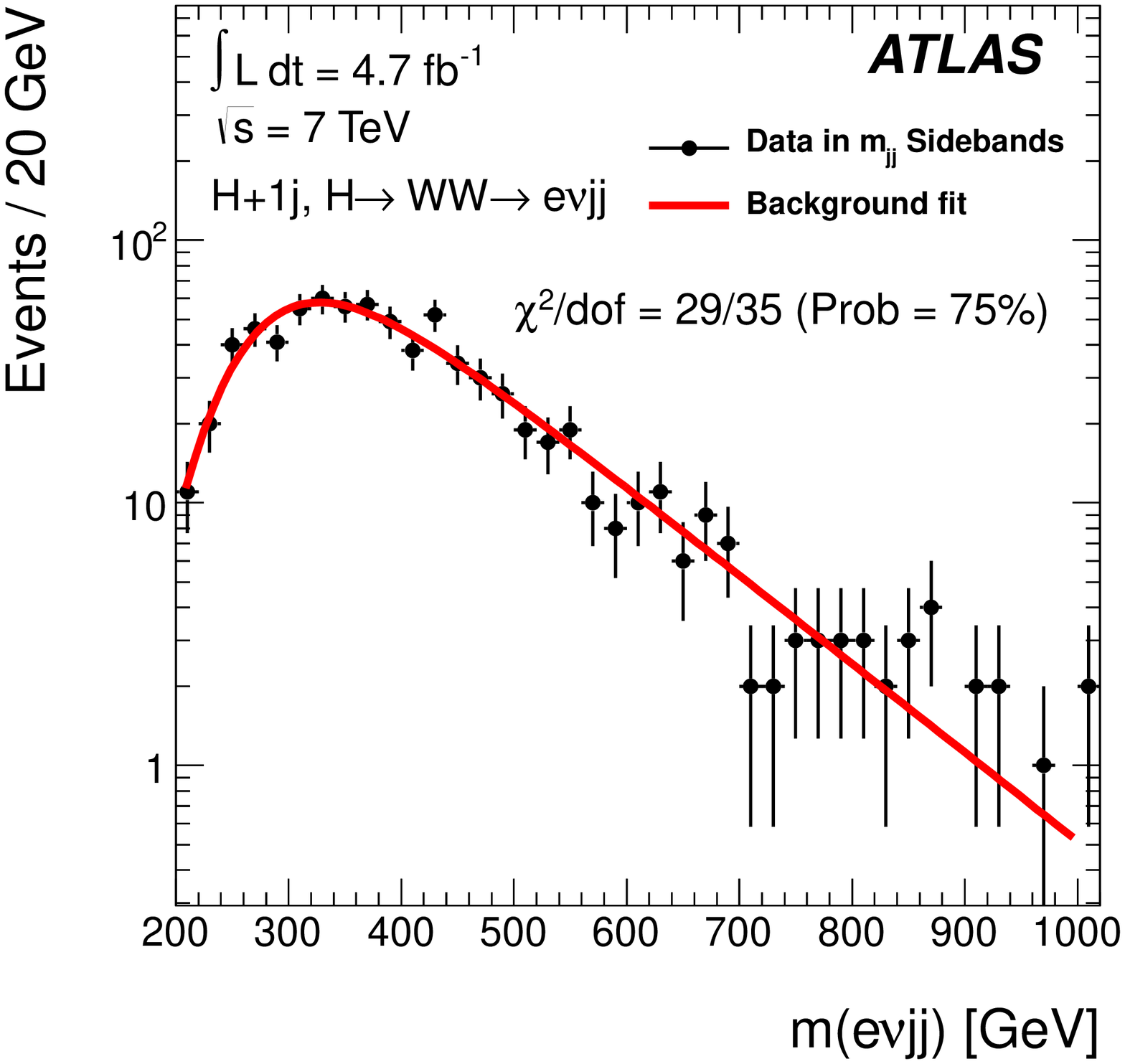} 
\includegraphics[angle=0,width=0.49\textwidth]{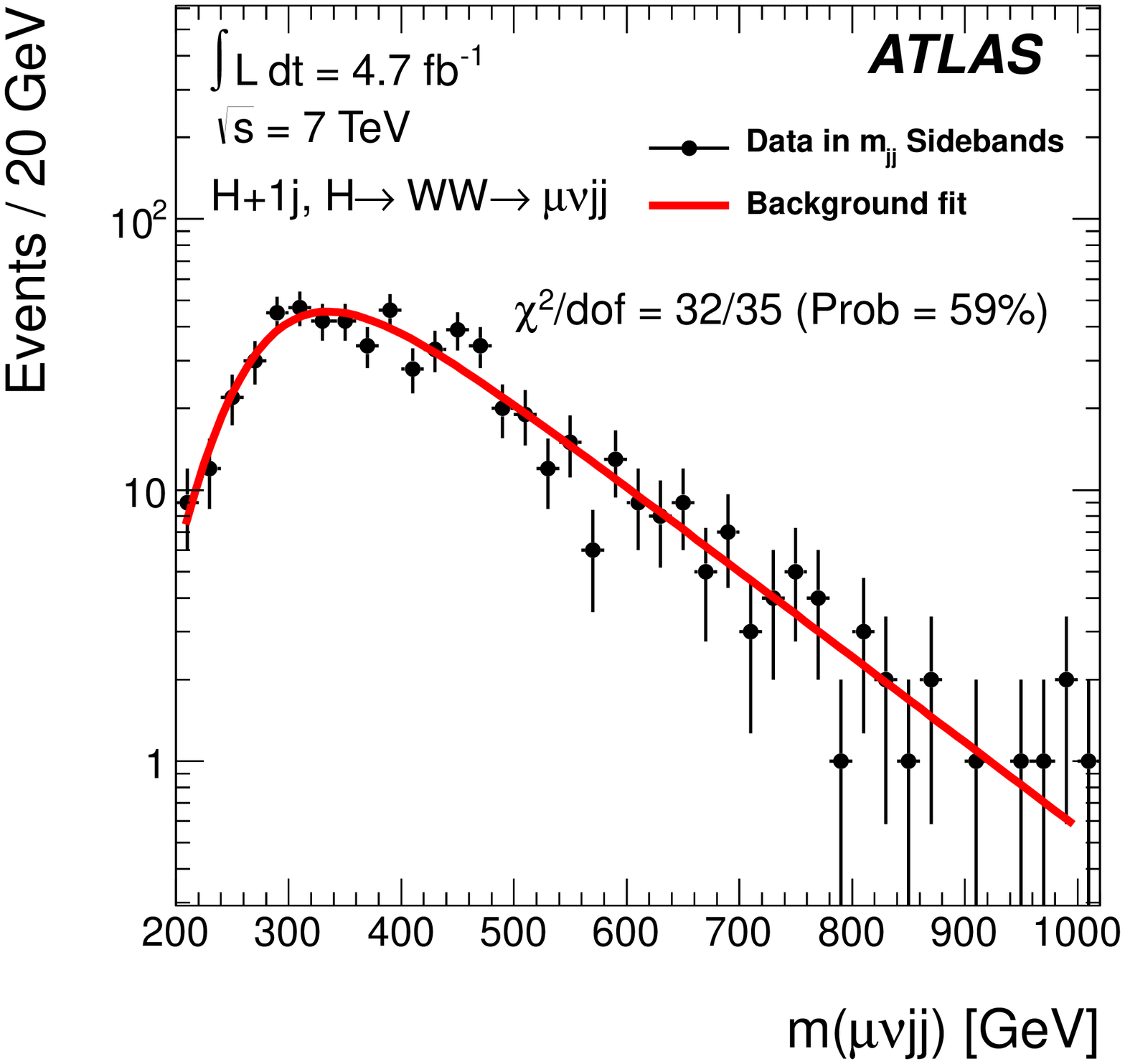}
\caption{Fits of the background model described in the text to the
  reconstructed invariant mass $m(\ell\nu jj)$ when the $m_{jj}$ is in
  the $W$ sidebands for the $\ell\nu jj+1j$ selection. The left (right) figure
  shows the electron (muon) channel distributions. The $\chi^2$/dof
  and $\chi^2$ probability of these fits are also shown in the figure.
}
\label{sideband_fits_1j}
\end{center}
\end{figure*}

The Higgs boson signal is expected to appear as a peak in the
$m(\ell\nu jj)$ distribution. Its width, before detector effects,
varies from about 10~GeV at $m_H = 300$ GeV to about 70~GeV at $m_H =
550$ GeV. The non-resonant background for the $\ell\nu jj+0/1j$
channel is modelled by a smooth function of the form $f(x) =
[1/(1+|a(x-m)|^b)] \times {\rm exp}[-c(x-200)]$, where $x$ is
$m(\ell\nu jj)$ in GeV and $a$, $b$, $c$, and $m$ are free parameters
with the appropriate units. In the $\ell\nu jj+2j$ channel, the
background is modelled by the sum of two exponential functions. The
parameters of the fitted function in each of these models are not
subjected to any external constraint. The functional form for the
background model is well motivated by studies using MC simulation, and
is tested by fits to the $m(\ell\nu jj)$ distributions obtained
through event selection in the $W$ sidebands, with $m_{jj}$ just below
($45~\GeV < m_{jj} < 60~\GeV$) or just above ($100~\GeV < m_{jj} <
115~\GeV$) the $W$ boson peak. Figures~\ref{sideband_fits_0j} and
\ref{sideband_fits_1j} show fits of the $\ell\nu jj$ mass to the
background model for $\ell\nu jj+0j$ and $\ell\nu jj+1j$ selections
with $m_{jj}$ in the $W$ sidebands. The ~$\chi^2$ probabilities of
these fits are between 25\% and 75\%, providing support for the
background functional form used in this analysis.

MC simulation is used to study the expected Higgs boson contribution 
to the $m(\ell\nu jj)$ distributions. Both the gluon fusion and the
weak boson fusion signal production processes are simulated using
the POWHEG~\cite{powheg,Nason:2009ai} event generator interfaced to 
PYTHIA~\cite{pythia} and are normalized to the next-to-next-to-leading
order cross sections~\cite{LHCHiggsCrossSectionWorkingGroup:2012vm}
shown in Table~\ref{signalCrossSectionTable}.
\begin{table*}
\begin{center}
\caption{Cross sections for Standard Model Higgs boson production and
  the branching ratio (BR) for $H\rightarrow WW\rightarrow\ell\nu jj$
  ($\ell=e$ or $\mu$) as a function of Higgs boson mass $\mH$. 
  The cross section and its associated uncertainties are described in
  Ref.~\cite{LHCHiggsCrossSectionWorkingGroup:2011ti}. The branching
  ratio includes $W\to\tau\to\ell$, and the uncertainties from the
  subchannels~\cite{Nakamura:2010zzi} are added in quadrature with the
  $H\to WW$ uncertainty, which is 0.5\% below 500~GeV and
  $0.1 m_H^4$ for $m_H\gtrsim500$~GeV.
}
\label{signalCrossSectionTable}
\begin{tabular}{ c | c c | c }
\hline
\hline
$\mH$[GeV]		     & $\sigma(gg\rightarrow H)$ [pb]   & $\sigma(qq\rightarrow H)$ [pb] & BR($H\rightarrow\ell^{\pm}\nu jj$)\\
\hline
300\rule[-1mm]{0mm}{4.7mm}   & 2.4~$\pm$~0.4                    & $0.30\pm0.01$                   & $0.237\pm0.003$\\  
400\rule[-1mm]{0mm}{4.7mm}   & 2.0~$\pm$~0.3			& 0.162$^{+0.010}_{-0.005}$        & $0.199\pm0.002$\\  
500\rule[-1mm]{0mm}{4.7mm}   & 0.85~$\pm$~0.15			& 0.095$^{+0.007}_{-0.003}$        & $0.187\pm0.002$\\  
600\rule[-2mm]{0mm}{5.7mm}   & 0.33~$\pm$~0.06		        & 0.058$^{+0.005}_{-0.002}$        & $0.191\pm0.003$\\  
\hline
\hline
\end{tabular}
\end{center}
\end{table*}
The $m(\ell\nu jj)$ distribution for the expected signal at each
hypothesized $m_H$ is modelled using the functional form
$1/(a+(x-m_1)^2+b(x-m_2)^4)$ with parameters ($a$, $b$, $m_1$ and
$m_2$) determined from a fit to the MC simulation of the expected
Higgs boson signal. The $m(l \nu jj)$ fractional resolution is $8.8
\pm 1.3$\% at $\mH=400$ GeV, the uncertainty arising mostly from the
\met\ and jet energy scale as described below, and shows a
$1/\sqrt{\mH}$ dependence over the range of this analysis.

\section{Systematic Uncertainties}

The systematic uncertainty due to the background modelling is included by 
treating the uncertainties on the background model parameters
resulting from fits to the data as nuisance parameters in the
statistical interpretation of the data. Both the background model and
the sum of signal and background models are found to be good fits to
the data. For $m_H = 400$ GeV, the $\chi^2$ probabilities are 33\% and
31\% for the background-only and background-plus-signal fits,
respectively. Therefore, alternative parameterizations of the
background expectation that are consistent with the data will also be
consistent with the background model within its uncertainties. This is
tested by fitting both the signal region and the sideband regions of
the data with two alternative parameterizations. Differences in the
fitted background yield between these parameterizations and the
nominal background model are less than 5\%, while the uncertainty from
the nuisance parameters and statistical uncertainty is 10-12\%.

The remaining systematic uncertainties are related to the Higgs boson
signal. The fit includes nuisance parameters which account for the
uncertainty in the reconstruction efficiency. The trigger
efficiencies, the electron and muon reconstruction efficiencies,
lepton energy resolution and scale are varied within their
uncertainties, giving an uncertainty in the signal efficiency of less
than 1\%. Varying the jet energy scale~\cite{JES} within its
uncertainties yields an uncertainty of up to 8\% in the expected
signal in the $\ell\nu jj+0/1j$ channel for $m_H\geqslant
400$~GeV. Smearing the jet energies within the uncertainty on their
resolutions~\cite{ATLAS-CONF-2010-054} results in a signal uncertainty 
of 7\% for $m_H=400$~GeV and 5\% for $m_H=600$~GeV. The reconstructed
\met\ ~\cite{MET} is also affected by the uncertainties on the energy
scales and resolutions of reconstructed leptons and jets. The signal
uncertainties given above include the propagation of these effects 
to the reconstructed \met. The propagation to \met\ adds a
small contribution to the overall signal uncertainty. In addition, a
7\% uncertainty on the degradation of the \met\ resolution and scale
due to pile-up effects is estimated, which results in a negligible
uncertainty on the signal efficiency. The looser selection criteria for
the $\ell\nu jj+2j$ channel result in an 11\% uncertainty on the signal
efficiency from the jet energy scale at $m_H=400$~GeV while the 
uncertainty due to the jet energy resolution is 16\%.
The uncertainty on the $b$-tagging efficiency~\cite{ATLAS-btag-calib}
gives a maximum uncertainty of 8\% on the signal efficiency and shows
no strong dependence on $m_H$ or the selection criteria. 

The uncertainties on jet energy resolution and jet energy scale, which
also have an impact on \met, lead to systematic uncertainties on the
Higgs boson mass resolution (5\%) and on the Higgs boson mass scale
(2\%). These uncertainties are not included since their effect on the
fitted Higgs boson yield is considerably smaller than the systematic
uncertainty on the signal acceptance due to jet energy scale and
resolution. 

The Higgs boson signal expectation includes a 3.9\% systematic
uncertainty due the luminosity
determination~\cite{lumipaper,ATLAS-CONF-2011-116} and a
19.4\% uncertainty on the predicted Higgs boson cross
section~\cite{LHCHiggsCrossSectionWorkingGroup:2012vm}, taken to be
independent of the mass. Off-shell effects and interference between
the signal and background processes are discussed in
Refs.~\cite{LHCHiggsCrossSectionWorkingGroup:2012vm,Anastasiou:2011pi,Campbell:2011cu}. To
account for the uncertainties from these effects, an uncertainty of
150\%$\times m_{H}^{3}$ ($m_{H}$ in TeV) on the signal cross section
is included in the statistical interpretation of the data, where the
$m_{H}^{3}$ form is motivated by the scaling of the Higgs boson width
with $m_{H}$ and the normalization factor of 150\% is chosen to give
$\sim$30\% at $m_{H}=600$~GeV~\cite{LHCHiggsCrossSectionWorkingGroup:2012vm}.

\section{Results and Conclusions}

\begin{figure*}
\begin{center}
\includegraphics[angle=0,width=0.47\textwidth]{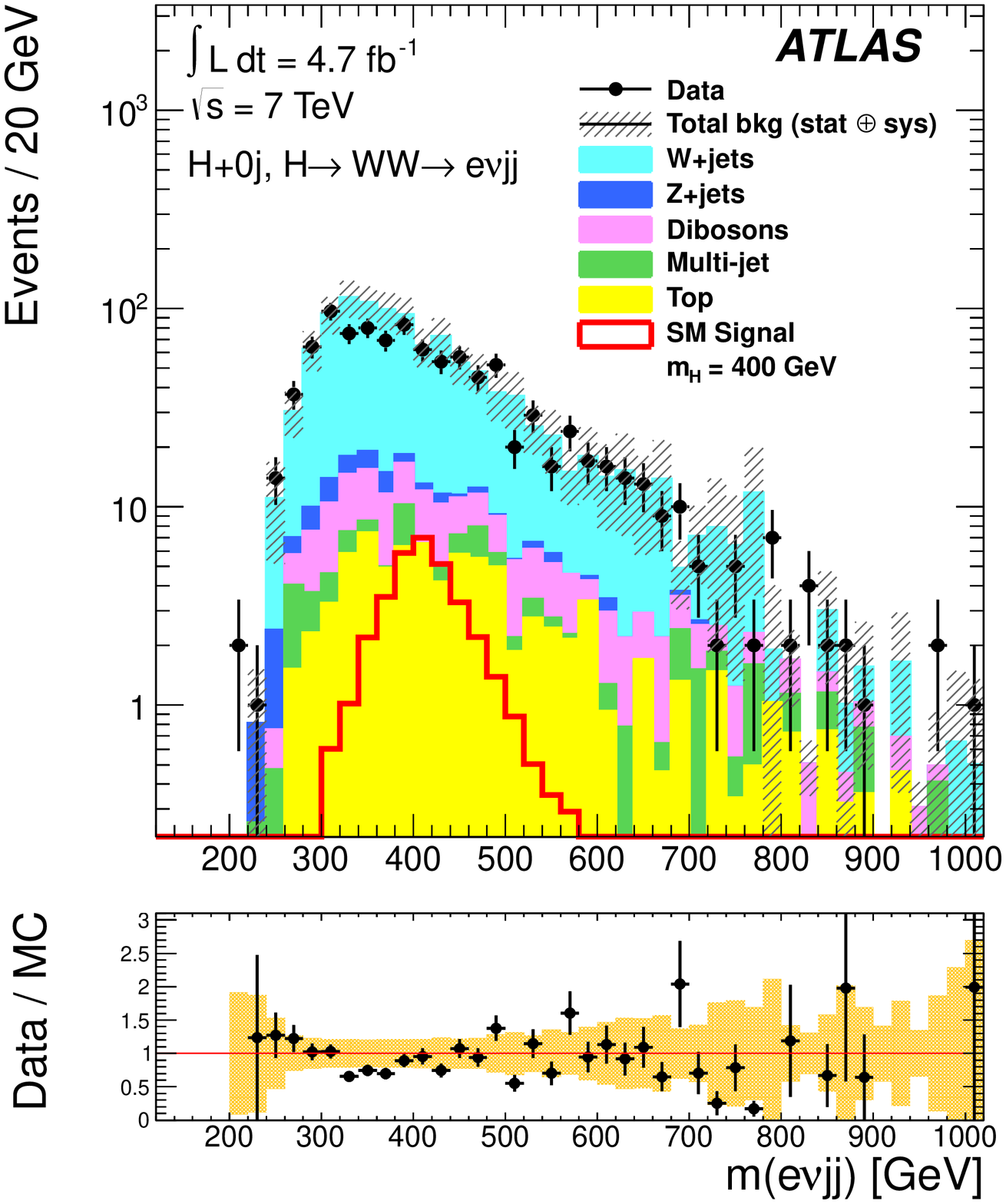} 
\includegraphics[angle=0,width=0.47\textwidth]{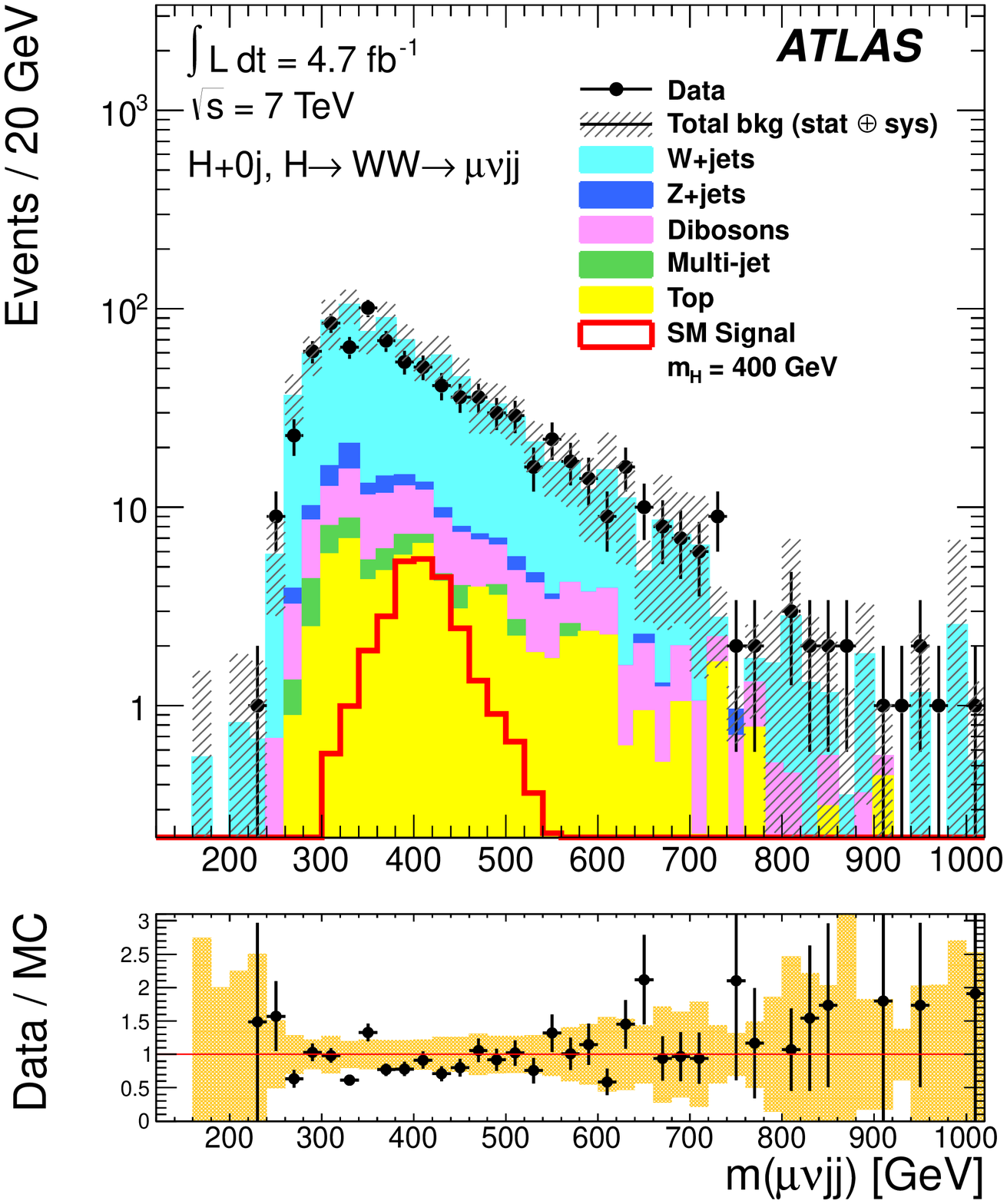}
\caption{
The reconstructed invariant mass $m(\ell\nu jj)$ in the data and expected
backgrounds using MC simulation for the $\ell\nu jj+0j$
selection. The left (right) figure shows the electron (muon) channel
distribution. The expected Higgs boson signal for $\mH = 400$ GeV is
also shown. The bottom panels show the data divided by the MC
expectation as markers, and the shaded (orange) region indicates the
systematic uncertainty on the background expectation from MC simulation.}
\label{mlnuqq0j_plots}
\end{center}
\end{figure*}

\begin{figure*}
\begin{center}
\includegraphics[angle=0,width=0.47\textwidth]{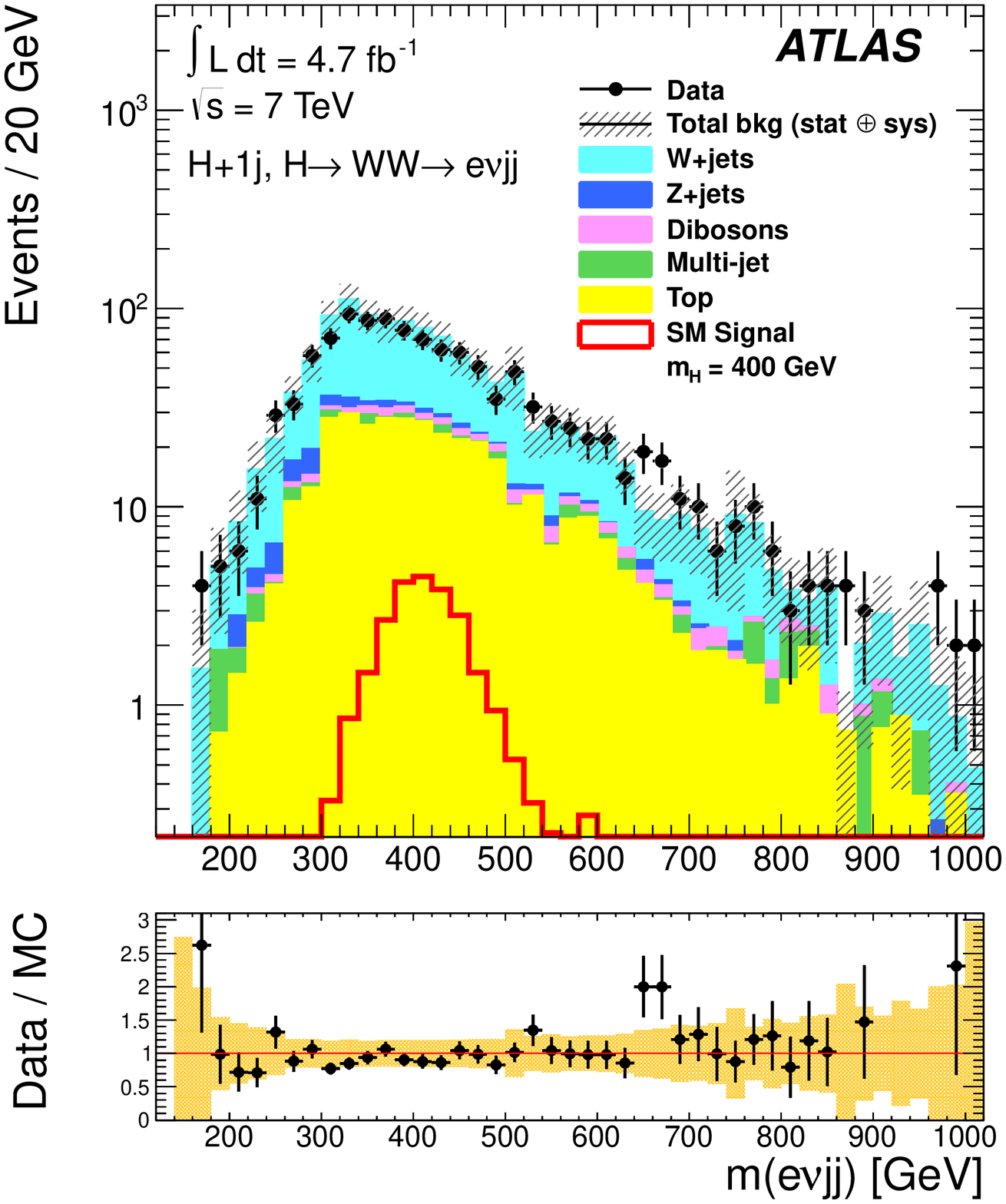} 
\includegraphics[angle=0,width=0.47\textwidth]{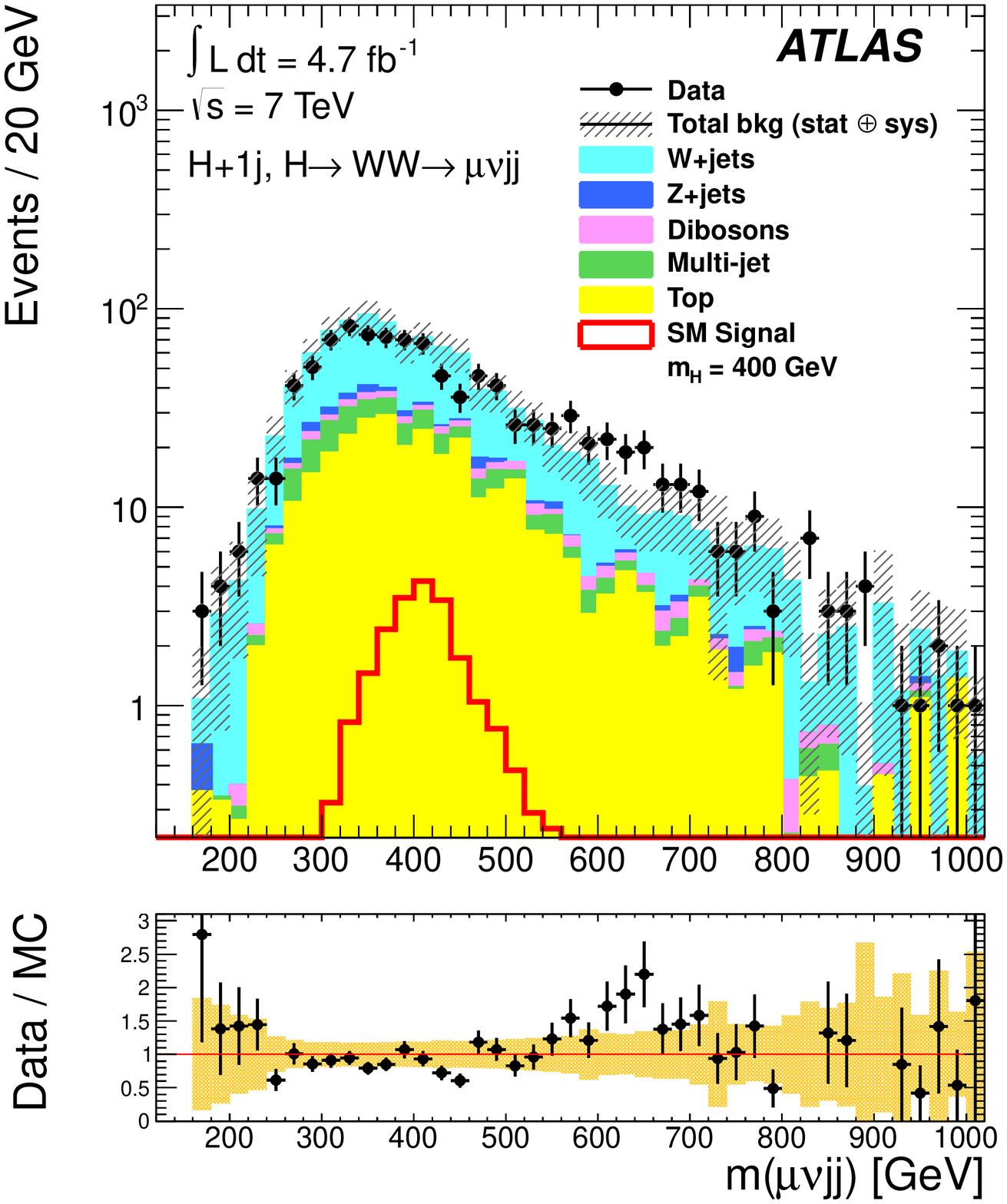}
\caption{
The reconstructed invariant mass $m(\ell\nu jj)$ in the data and expected
backgrounds using MC simulation for the $\ell\nu jj+1j$
selection. The left (right) figure shows the electron (muon) channel
distribution. The expected Higgs boson signal for $\mH = 400$ GeV is
also shown. The bottom panels show the data divided by the MC
expectation as markers, and the shaded (orange) region indicates the
systematic uncertainty on the background expectation from MC simulation.} 
\label{mlnuqq1j_plots}
\end{center}
\end{figure*}

\begin{figure*}
\begin{center}
\includegraphics[angle=0,width=0.47\textwidth]{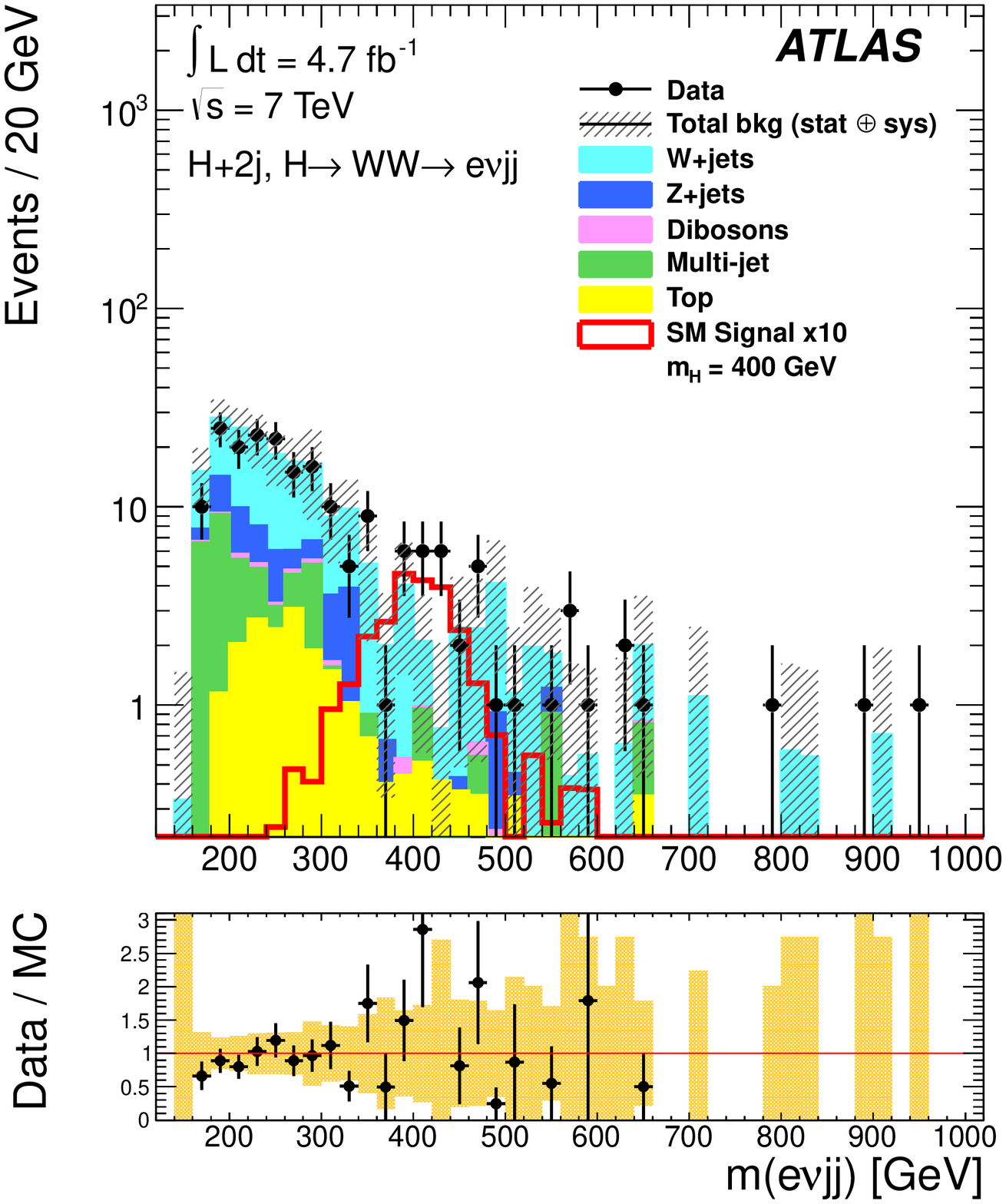} 
\includegraphics[angle=0,width=0.47\textwidth]{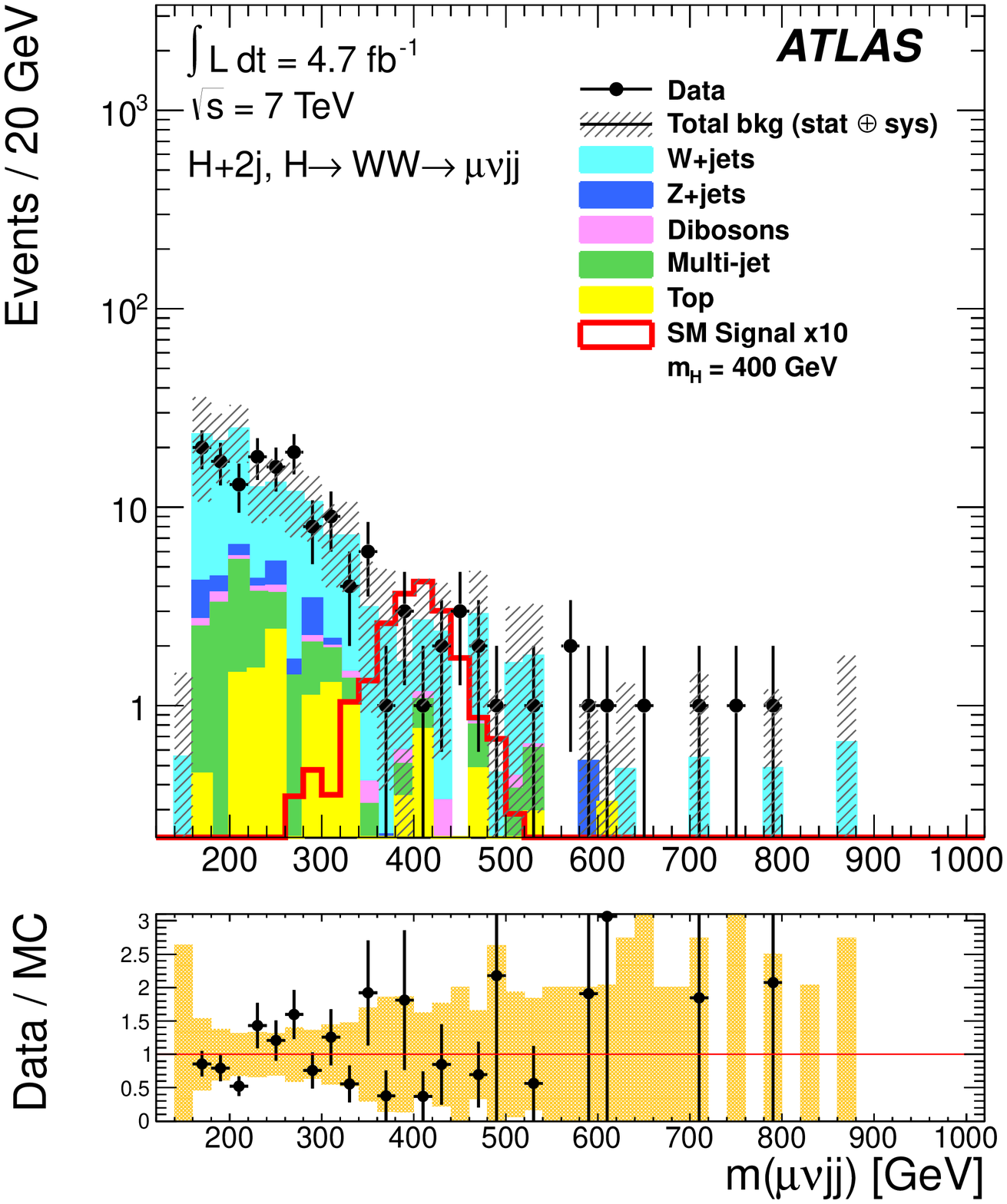}
\caption{
The reconstructed invariant mass $m(\ell\nu jj)$ in the data and expected
backgrounds using MC simulation for the $\ell\nu jj+2j$
selection. The left (right) figure shows the electron (muon) channel
distribution. The expected Higgs boson signal for $\mH =  400$~GeV is
also shown, scaled up by a factor of 10 for visibility. The bottom
panels show the data divided by the MC expectation as markers, and the
shaded (orange) region indicates the systematic uncertainty on the background
expectation from MC simulation.}
\label{mlnuqq2j_plots}
\end{center}
\end{figure*}

Figures~\ref{mlnuqq0j_plots}--\ref{mlnuqq2j_plots} show the $m(\ell\nu
jj)$ distributions and the ratio of data to background expectation
from MC simulation for the six different final states considered in
this analysis, along with bands showing the total background
uncertainty. The simulated background is not used in the statistical
interpretation of the data. Instead, the parameterizations described
in Section~\ref{sec:modeling} are used to model the background.

The Higgs boson signal yield in each final state is determined using a
binned maximum likelihood fit to the observed $m(\ell\nu jj)$
distribution in the range $200~\GeV < m(\ell\nu jj) < 2000~\GeV$. As a
check, fits over a smaller range ($200~\GeV < m(\ell\nu jj) <
1000~\GeV$) were also performed and the results were found to be
consistent with the results presented here.

\begin{figure*}
\begin{center}
\includegraphics[angle=0,width=0.49\textwidth]{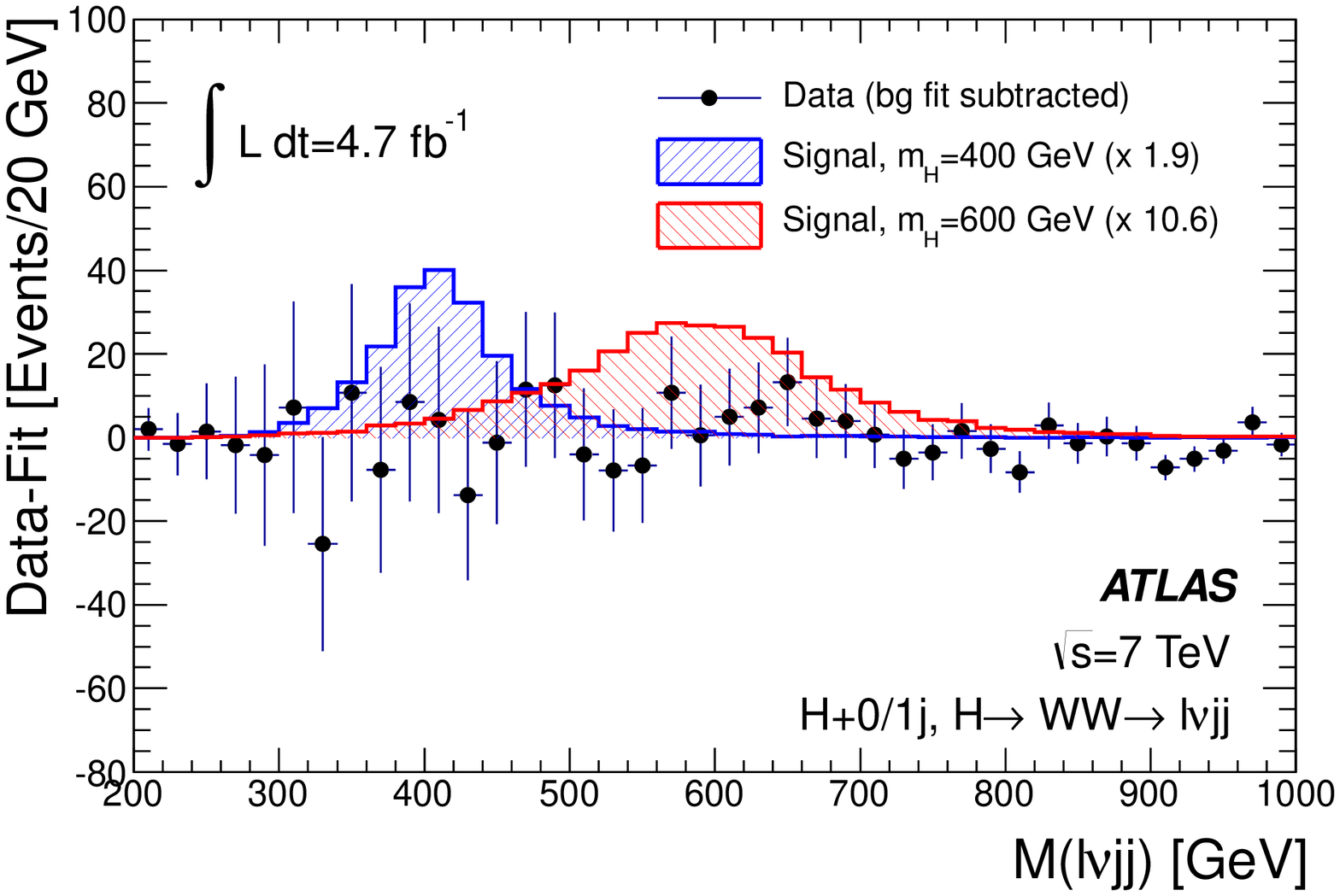} 
\includegraphics[angle=0,width=0.49\textwidth]{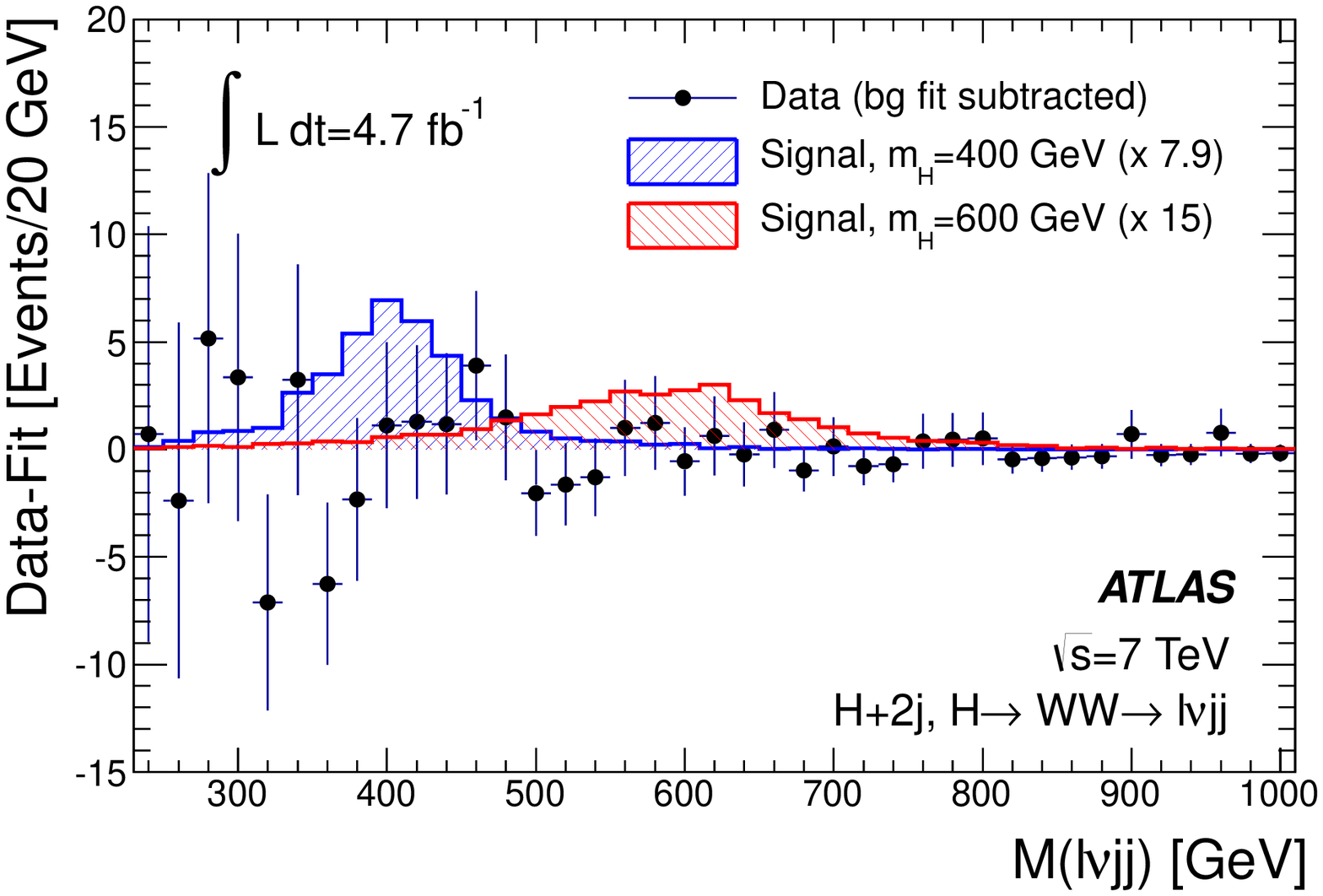}
\caption{
The difference between data and the fitted background under a
no-signal hypothesis, for the (left) $\ell\nu jj+0/1j$ selection
and (right) $\ell\nu jj+2j$ selection, both summed over lepton
flavours. The expected contribution from SM Higgs boson decays is also
shown for $\mH = 400$ GeV and $\mH = 600$ GeV, multiplied by
a factor equal to the ratio of 95\% CL limit on its production
to the SM prediction.
}
\label{sub_mlnuqq01j_plots}
\end{center}
\end{figure*}

The difference between data and the fitted background is shown in
Figure~\ref{sub_mlnuqq01j_plots}. The expected signals for $\mH =
400~\GeV$ and $\mH = 600~\GeV$ are also shown, each scaled to the 95\%
CL limit on the production cross section.

Figure~\ref{sub_mlnuqq01j_plots} shows that there is no indication
of a significant excess of data above the background model. Limits on
SM Higgs boson production are extracted using the profile
likelihood ratio~\cite{asymptotics} as a test statistic and following
the $CL_{s}$ procedure described in
Refs.~\cite{Read:2002hq,ATLAS:2012ae}.

\begin{figure*}[tbp]
\begin{center}
\includegraphics[angle=0,width=0.49\textwidth]{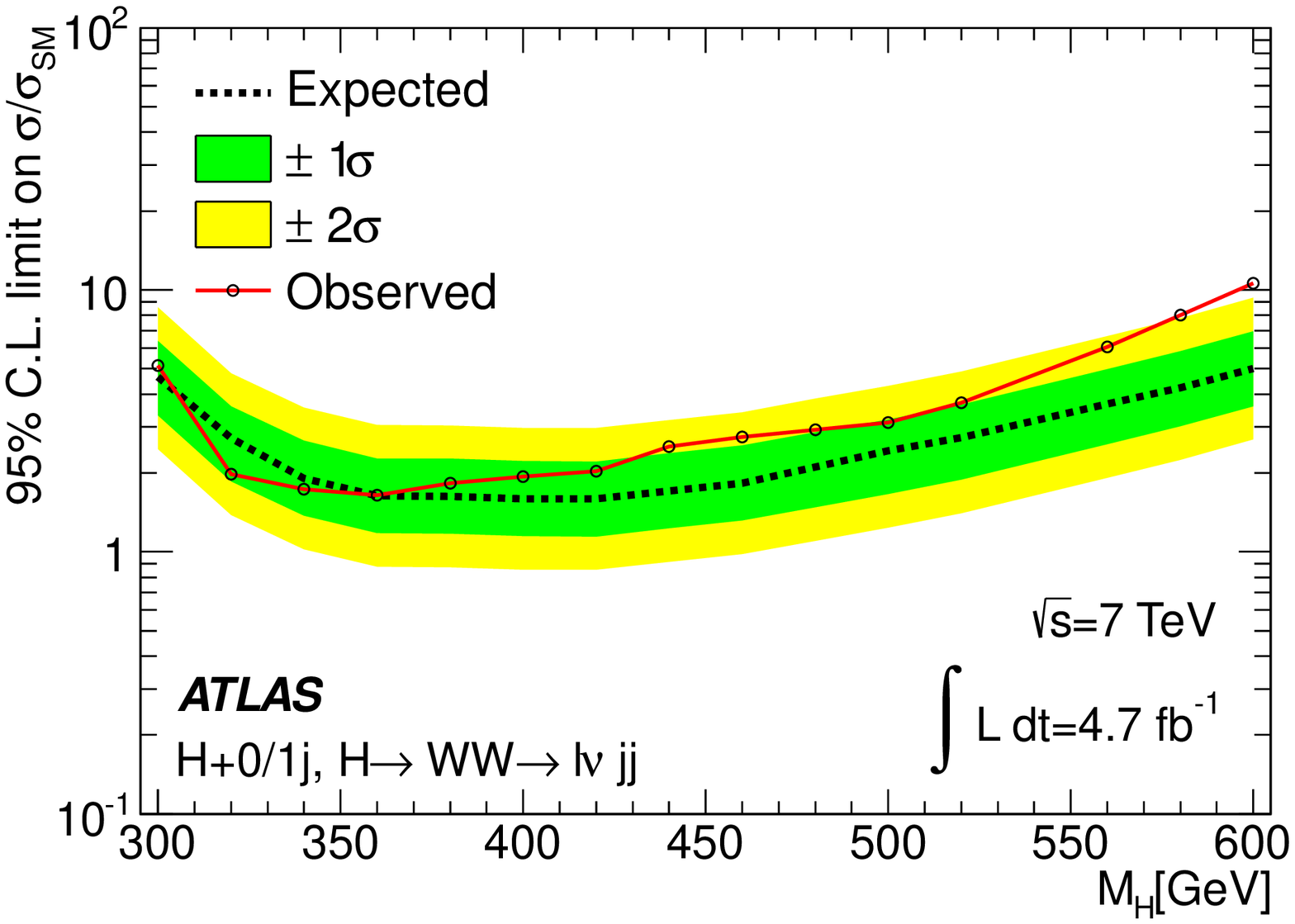}
\includegraphics[angle=0,width=0.49\textwidth]{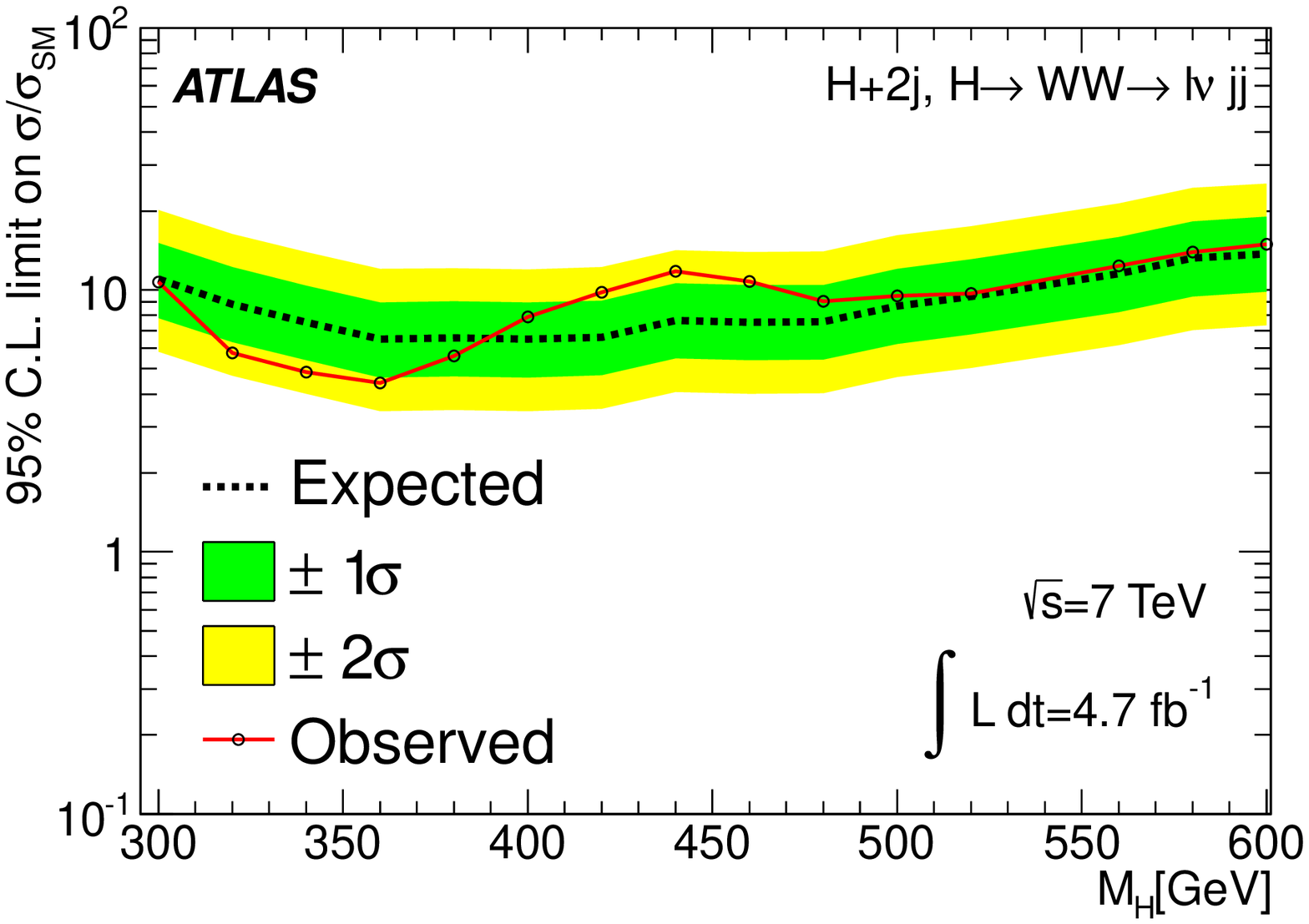}
\caption{
The expected and observed 95\% CL upper limits on the Higgs boson
production cross section divided by the SM prediction. The left figure
shows the combination of $H+0j$ with $H+1j$ and the right figure shows the
$H+2j$ limits. For any hypothesized Higgs boson mass, the background
contribution used in the calculation of this limit is obtained from a
fit to the $m(\ell\nu jj)$ distribution. The dark (green) and light
(yellow) bands show the $\pm 1\sigma$ and $\pm 2\sigma$ uncertainties
on the expected limit.
}
\label{limit_plot}
\end{center}
\end{figure*}

\begin{figure*}[tbp]
\begin{center}
\includegraphics[angle=0,width=0.80\textwidth]{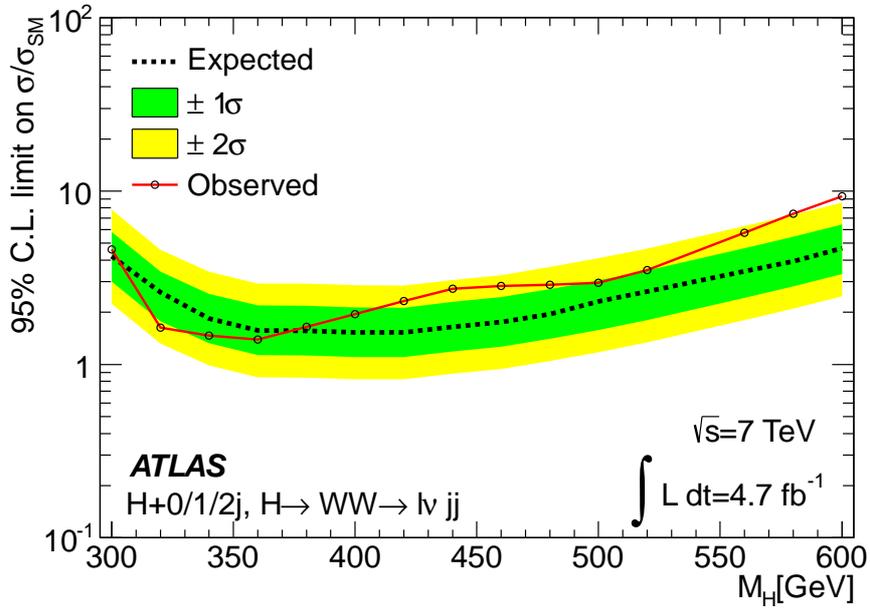}
\caption{
The expected and observed 95\% CL upper limits on the Higgs boson
production cross section divided by the SM prediction. This figure
shows the combination of the $H+0j$, $H+1j$ and $H+2j$ channels. The
background contribution used in the calculation of this limit is
obtained from a fit to the $m(\ell\nu jj)$ distribution. The green and
yellow bands show the $\pm 1\sigma$ and $\pm 2\sigma$ uncertainties on
the expected limit.
}
\label{limit_plot_012j}
\end{center}
\end{figure*}

\begin{figure*}[tbp]
\begin{center}
\includegraphics[angle=0,width=0.49\textwidth]{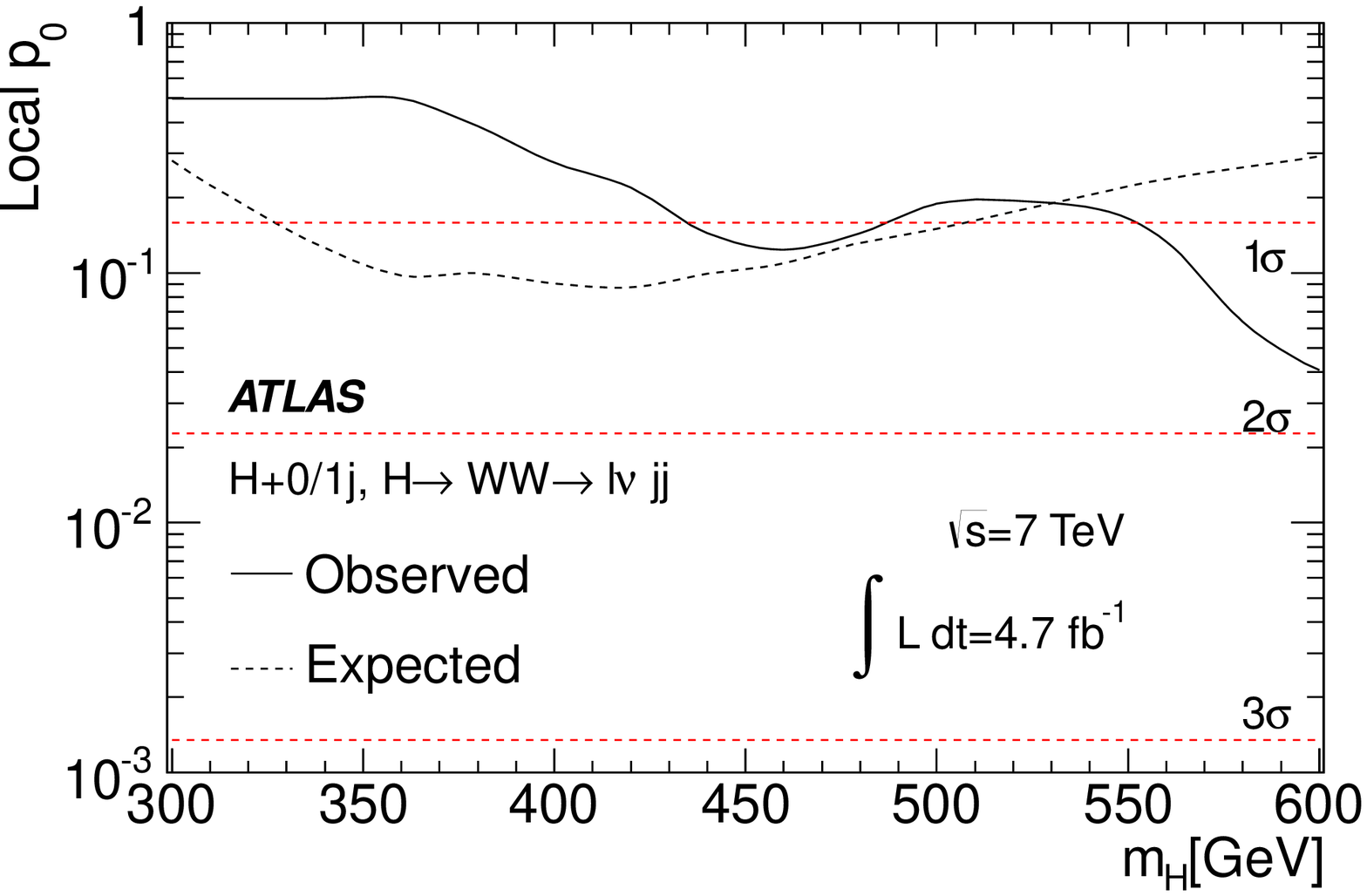}
\includegraphics[angle=0,width=0.49\textwidth]{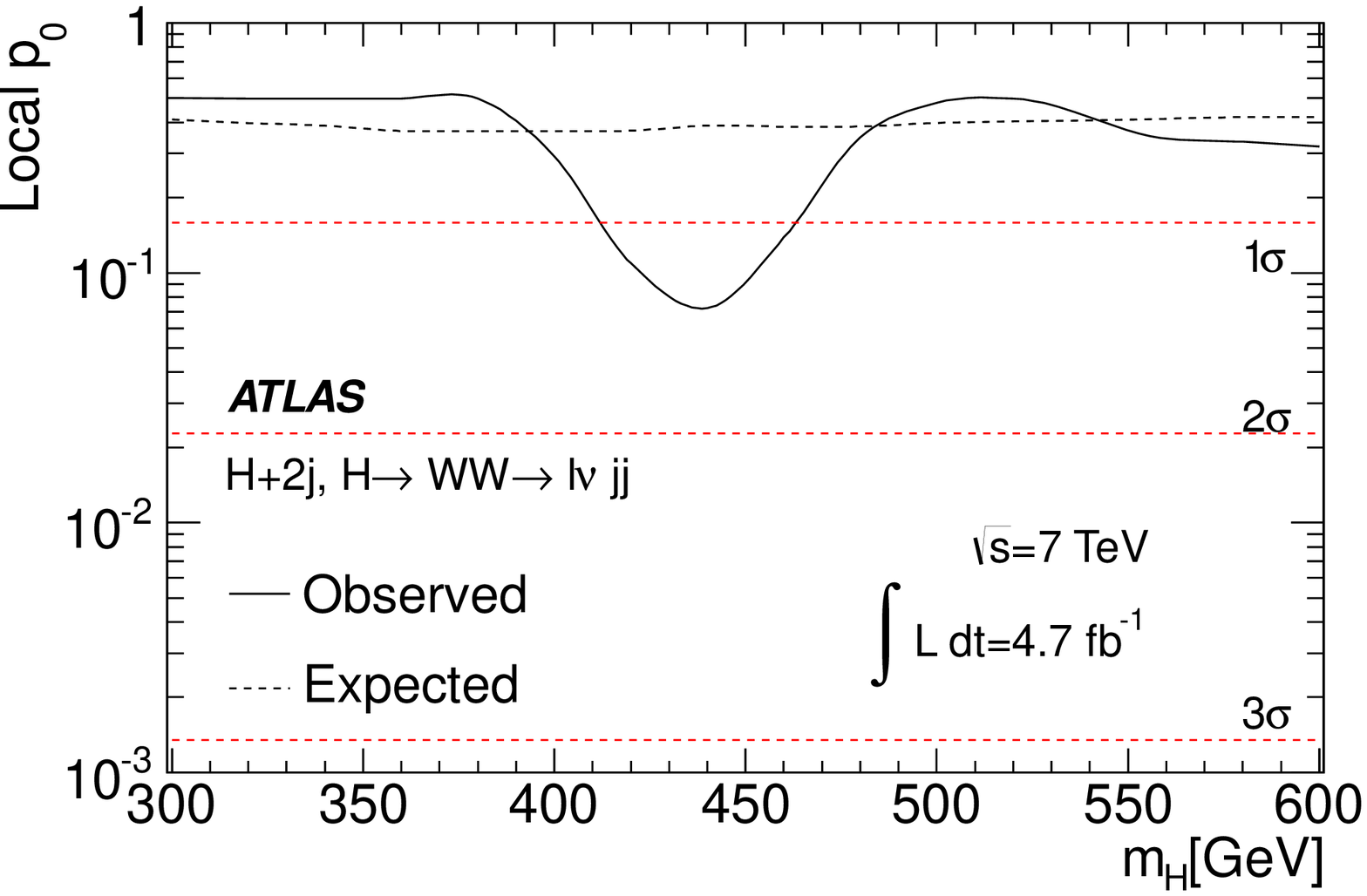}
\caption{Local $p_{0}$ for the SM Higgs boson search in the $H+0/1j$
  channel (left) and $H+2j$ channel (right). The dashed line shows the
  expected $p_0$ value for a Standard Model Higgs boson as a function
  of its mass.
\label{pvalues}
}
\end{center}
\end{figure*}

Figure~\ref{limit_plot} shows the 95\% CL upper bound on the cross
section times branching ratio for Higgs boson production with respect
to the Standard Model prediction, as a function of $\mH$. The best
sensitivity is reached at $\mH = 400~\GeV$, where the 95\% confidence
level upper bound on the cross section for $H\rightarrow WW$
production using the combined $H+0j$ and $H+1j$ channels is observed
(expected) to be 2.2 pb (1.9 pb) corresponding to 1.9 (1.6) times the
Standard Model prediction. In the $H+2j$ channel, which is more
sensitive to Higgs boson production via weak boson fusion, the 95\%
confidence level upper bound on the cross section for $H\rightarrow
WW$ production with $\mH = 400\GeV$ is observed (expected) to be 0.7
pb (0.6 pb) corresponding to 7.9 (6.5) times the Standard Model
prediction. Figure \ref{limit_plot_012j} shows the limits obtained
when combining the $H+2j$ channel with the $H+0/1j$
channels. Figure~\ref{pvalues} shows the probability $p_{0}$ to
observe a fluctuation in $300 < m(\ell\nu jj) < 600~\GeV$ at least as
large as the one observed in data if there is no signal contribution,
where the signal and background are modelled as described in Section
\ref{sec:modeling}. The expected $p_{0}$ for $H+0/1j$ if there were a
SM Higgs at 400 GeV is 0.091, and the observed value is 0.276. For
$H+2j$, the expected $p_{0}$ is 0.369 and the observed is 0.293. The
significance is computed as $\sqrt{-2\log\lambda}$ where $\lambda$ is
the likelihood ratio obtained by the fit, and the significance is
converted into the probability $p_0$ using the Gauss error function.

In summary, a search for the SM Higgs boson has been performed in the
$\HWWlvqq$ channel using 4.7 $\ifb$ of $pp$ collisions at $\sqrt{s}$ =
7 TeV recorded by the ATLAS detector. No significant excess of events 
over the expected background has been observed. Exclusion limits on SM
Higgs boson production at 95\% CL are reported over the Higgs boson
mass range of $300-600$ GeV.


\section{Acknowledgements}

We thank CERN for the very successful operation of the LHC, as well as the
support staff from our institutions without whom ATLAS could not be
operated efficiently.

We acknowledge the support of ANPCyT, Argentina; YerPhI, Armenia; ARC,
Australia; BMWF, Austria; ANAS, Azerbaijan; SSTC, Belarus; CNPq and FAPESP,
Brazil; NSERC, NRC and CFI, Canada; CERN; CONICYT, Chile; CAS, MOST and NSFC,
China; COLCIENCIAS, Colombia; MSMT CR, MPO CR and VSC CR, Czech Republic;
DNRF, DNSRC and Lundbeck Foundation, Denmark; EPLANET and ERC, European Union;
IN2P3-CNRS, CEA-DSM/IRFU, France; GNAS, Georgia; BMBF, DFG, HGF, MPG and AvH
Foundation, Germany; GSRT, Greece; ISF, MINERVA, GIF, DIP and Benoziyo Center,
Israel; INFN, Italy; MEXT and JSPS, Japan; CNRST, Morocco; FOM and NWO,
Netherlands; RCN, Norway; MNiSW, Poland; GRICES and FCT, Portugal; MERYS
(MECTS), Romania; MES of Russia and ROSATOM, Russian Federation; JINR; MSTD,
Serbia; MSSR, Slovakia; ARRS and MVZT, Slovenia; DST/NRF, South Africa;
MICINN, Spain; SRC and Wallenberg Foundation, Sweden; SER, SNSF and Cantons of
Bern and Geneva, Switzerland; NSC, Taiwan; TAEK, Turkey; STFC, the Royal
Society and Leverhulme Trust, United Kingdom; DOE and NSF, United States of
America.

The crucial computing support from all WLCG partners is acknowledged
gratefully, in particular from CERN and the ATLAS Tier-1 facilities at
TRIUMF (Canada), NDGF (Denmark, Norway, Sweden), CC-IN2P3 (France),
KIT/GridKA (Germany), INFN-CNAF (Italy), NL-T1 (Netherlands), PIC (Spain),
ASGC (Taiwan), RAL (UK) and BNL (USA) and in the Tier-2 facilities
worldwide.

\bibliographystyle{atlasnote}
\bibliography{HWWlvqqPaper}

\clearpage 
\onecolumn
\begin{flushleft}
{\Large The ATLAS Collaboration}

\bigskip

G.~Aad$^{\rm 48}$,
B.~Abbott$^{\rm 111}$,
J.~Abdallah$^{\rm 11}$,
S.~Abdel~Khalek$^{\rm 115}$,
A.A.~Abdelalim$^{\rm 49}$,
O.~Abdinov$^{\rm 10}$,
R.~Aben$^{\rm 105}$,
B.~Abi$^{\rm 112}$,
M.~Abolins$^{\rm 88}$,
O.S.~AbouZeid$^{\rm 158}$
% ...
\bigskip

% ...
$^{10}$ Institute of Physics, Azerbaijan Academy of Sciences, Baku, Azerbaijan\\
$^{11}$ Institut de F\'isica d'Altes Energies and Departament de F\'isica de la Universitat Aut\`onoma  de Barcelona and ICREA, Barcelona, Spain\\
% ...
$^{48}$ Fakult\"{a}t f\"{u}r Mathematik und Physik, Albert-Ludwigs-Universit\"{a}t, Freiburg i.Br., Germany\\
$^{49}$ Section de Physique, Universit\'e de Gen\`eve, Geneva, Switzerland\\
% ...
$^{88}$ Department of Physics and Astronomy, Michigan State University, East Lansing MI, United States of America\\
% ...
$^{105}$ Nikhef National Institute for Subatomic Physics and University of Amsterdam, Amsterdam, Netherlands\\
% ...
$^{111}$ Homer L. Dodge Department of Physics and Astronomy, University of Oklahoma, Norman OK, United States of America\\
$^{112}$ Department of Physics, Oklahoma State University, Stillwater OK, United States of America\\
% ...
$^{115}$ LAL, Universit\'e Paris-Sud and CNRS/IN2P3, Orsay, France\\
% ...
$^{158}$ Department of Physics, University of Toronto, Toronto ON, Canada\\
% ...
\end{flushleft}


\end{document}
\bye